\documentclass[twocolumn,pra,showpacs,superscriptaddress]{revtex4}
\usepackage{amsmath}
\usepackage{graphicx}
\usepackage{amssymb}
\usepackage{amsmath}
\usepackage{graphicx}
\input{Qcircuit}

\newcommand{\NOT}{\textsc{not}}

\newcommand{\CNOT}{controlled-\NOT}

\newcommand{\sgate}[1]{\gate{\rule[-0.8ex]{0pt}{2.6ex}#1}}
\newcommand{\nuc}[2]{\mbox{${}^{#1}\rm #2$}}

\begin{document}
\title{Preparing pseudo-pure states with controlled-transfer gates}
\author{Minaru Kawamura}
\affiliation{Electrical and Electronic Engineering, Okayama University of Science, 1-1 Ridai-cho, Okayama, 700-0005 Japan}
\author{Benjamin Rowland}
\affiliation{Centre for Quantum Computation, Clarendon Laboratory, University of Oxford, Parks Road, OX1 3PU, United Kingdom}
\author{Jonathan A. Jones}\email{jonathan.jones@qubit.org}
\affiliation{Centre for Quantum Computation, Clarendon Laboratory, University of Oxford, Parks Road, OX1 3PU, United Kingdom}
\affiliation{Centre for Advanced ESR, University of Oxford, South Parks Road, Oxford OX1 3QR, United Kingdom}
\date{\today}
\begin{abstract}
The preparation of pseudo-pure states plays a central role in the implementation of quantum information processing in high temperature ensemble systems, such as nuclear magnetic resonance.  Here we describe a simple approach based on controlled-transfer gates which permits pseudo-pure states to be prepared efficiently using spatial averaging techniques.
\end{abstract}
\maketitle

Nuclear Magnetic Resonance (NMR) spin systems provide a simple approach for demonstrating techniques for quantum information processing \cite{Jones2001a}.  Because of the difficulty in preparing pure spin states almost all experiments have used \textit{pseudo-pure states} \cite{Cory1997,Cory1998a}, sometimes called \textit{effective pure states} \cite{Gershenfeld1997,Knill1998}.  These are mixed states of the form
\begin{equation}\label{eq:ppure}
(1-\delta)\frac{\mathbf{1}}{2^n}+\delta\ket{\textbf{0}}\bra{\textbf{0}}
\end{equation}
where $\ket{\textbf{0}}=\ket{00\dots0}$ is the desired ground state and $\mathbf{1}/2^n$ is the maximally mixed state.  The effective purity $\delta$ of these states is very low if prepared from high-temperature thermal systems \cite{Warren1997}, but the pseudo-pure state approach remains useful for simple demonstrations, and may also find applications in systems prepared at spin temperatures which are low enough to substantially enhance the ground state population but not low enough to generate a true pure state.

A range of methods for preparing pseudo-pure states have been described \cite{Jones2001a}, which can usually be divided into temporal averaging approaches, in which the final result is averaged over a number of repetitions of the experiment with different starting states, and spatial averaging approaches, in which magnetic field gradients are used to average the spin system over a macroscopic sample.  Spatial averaging has the advantage that only a single experiment is required, but suffers from a number of practical disadvantages: in particular it can be difficult to design experiments which prepare pseudo-pure states with the highest possible purity, and to avoid the generation of zero-quantum coherence terms \cite{EBWbook}, which are not averaged by field gradients.  Here we describe a simple spatial averaging approach which prepares pseudo-pure states with the highest possible purity.

\section{Controlled-transfer gates}
The thermal state of an NMR spin system is diagonal in the computational basis with a pattern of populations determined by Boltzmann factors.  Although details vary greatly, all methods for preparing pseudo pure states fundamentally work by averaging the populations of states other than the ground state, ideally leaving the ground state untouched \cite{Kawamura2004, Anwar2006}.

Averaging of this kind can be achieved using a \textit{controlled-transfer} gate, which comprises a controlled-rotation gate with a rotation angle of $\theta$, followed by the application of a magnetic field crush gradient which removes all off-diagonal terms (if the system's density matrix is diagonal before the gate zero-quantum terms will not be generated).  The controlled rotation occurs around some axis in the $xy$ plane; the choice of axis is unimportant as the off-diagonal terms, which depend on the choice of axis, are removed by the field gradient.  For definiteness the rotation can be thought of as occurring around the $x$-axis, so that a rotation with $\theta=\pi$ corresponds to a \CNOT\ gate.  In the remainder of this paper we use the network symbol
\begin{equation}
\mbox{
\Qcircuit @C=1em @R=.7em {
& \ctrl{1} & \qw \\
& \sgate{\theta}& \qw
}}
\end{equation}
to denote a controlled-transfer gate, with the crush gradient being implicit.  Related ideas have been explored using line-selective pulses \cite{Peng2001,Das2003a}.

The action of a controlled-transfer gate on a general population state of a two-qubit system is
\begin{equation}
\begin{pmatrix}a&0&0&0\\0&b&0&0\\0&0&c&0\\0&0&0&d\end{pmatrix}\longrightarrow \begin{pmatrix}a&0&0&0\\0&b&0&0\\0&0&c'&0\\0&0&0&d'\end{pmatrix}\label{eq:rhodiag}
\end{equation}
with
\begin{equation}
c'=\frac{c+d+(c-d)\cos\theta}{2}
\end{equation}
\begin{equation}
d'=\frac{c+d-(c-d)\cos\theta}{2}
\end{equation}
so that the populations of two states are partially averaged, with the difference in the two populations being scaled down by $\cos\theta$.  Interchanging the roles of the control and target qubits serves to average populations $b$ and $d$, and so the network
\begin{equation}
\mbox{
\Qcircuit @C=1em @R=.7em {
& \ctrl{1} & \sgate{\theta_2} & \qw \\
& \sgate{\theta_1}& \ctrl{-1} & \qw
}}\label{eq:network2}
\end{equation}
will mix the populations $b$, $c$ and $d$, while leaving $a$, the population of the ground state, untouched.  The final state will depend on the choice of rotation angles and the initial state; to generate a pseudo-pure state the angles must be set to
\begin{equation}
\theta_1=\arccos\left[\frac{2b-(c+d)}{3(c-d)}\right]\qquad\theta_2=\frac{\pi}{2}\label{eq:theta2spin}
\end{equation}
where the gates in Eq.~\ref{eq:network2} are applied from left to right, and the initial density matrix, Eq.~\ref{eq:rhodiag}, is defined with the state of the lowest qubit varying most rapidly.  If the order of the two gates is interchanged to give the network
\begin{equation}
\mbox{
\Qcircuit @C=1em @R=.7em {
& \sgate{\theta_1}&\ctrl{1} & \qw \\
& \ctrl{-1}&\sgate{\theta_2}& \qw
}}\label{eq:network2a}
\end{equation}
then the values of $b$ and $c$ must be interchanged in Eq.~\ref{eq:theta2spin}.

For a thermal state of a homonuclear spin system, so that the polarizations of the two nuclei are the same, the traceless part of the density matrix, sometimes called the deviation density matrix \cite{Chuang1998b}, is proportional to $\sigma^1_z+\sigma^2_z$. In this case $d=-a$ and $b=c=0$ so $\theta_1=\arccos(1/3)\approx70.5^\circ$.  As the desired state $\ket{00}$ has the largest population of any spin state at thermal equilibrium, and the controlled-transfer gates leave this population term untouched, this approach will automatically generate the desired pseudo-pure state with the largest effective purity which is possible by \textit{any} averaging process.  Methods based on exhaustive temporal averaging \cite{Knill1998} will also reach this limit, but the standard methods based on spatial averaging will not achieve this.

Note that as controlled-transfer gates only act to transfer population terms, and these transfers can be calculated analytically, there is no need to perform a full simulation of the whole density matrix.  Although this point is not particularly important in two-qubit systems, it becomes important in larger systems as the size of the density matrix rises very rapidly with the number of qubits.  This makes a full simulation impractical, while calculating the effects of controlled-transfer gates remains relatively tractable.

\section{Heteronuclear spin systems}
The method of controlled-transfer gates makes no assumptions about the initial density matrix beyond the fact that it is diagonal, and so can be used with any initial population state.  In heteronuclear spin systems, or spin-systems not starting at thermal equilibrium, it is of course necessary to use the more general formula in Eq.~\ref{eq:theta2spin}.  In some cases only one of the two networks, Eqns.~\ref{eq:network2} and~\ref{eq:network2a}, can be used, with the other network requiring an impossible value for $\theta_1$.  For heteronuclear spin systems starting at thermal equilibrium a solution can always be found if the target of the first controlled-transfer gate is the spin with the larger polarization; the angle of the first gate then rises from $\theta_1=\arccos(1/3)\approx70.5^\circ$ to $\theta_1=\pi-\arccos(1/3)\approx109.5^\circ$ as the polarization ratio is increased from unity towards infinity.

In heteronuclear spin systems the controlled-transfer approach gate approach can give even larger improvements in effective purity compared with the conventional spatial averaging techniques, firstly because some conventional spatial averaging methods \cite{Pravia1999} trade simplicity for efficiency, and secondly because such methods usually begin with an initial sequence that equalizes the polarization of the two spins.  Although an efficient method for doing this in a two spin system is known \cite{Pravia1999}, the simplest general method in larger spin systems is just to reduce the polarization of the more highly polarized spins to that of the least polarized spins, inevitably wasting polarization in the process.

A special case occurs when the polarization of one of the spins is twice that of the other, as in this case a pseudo-pure state can be prepared with a single controlled-transfer gate with an angle of $\pi/2$ applied to the more polarized spin.  This individual result has been known for may years, and has applied to homonuclear spin systems where the polarization of one spin has been reduced \cite{Cory1998a} and to heteronuclear spin systems where the excess polarization of the more polarized spin has been reduced to a factor of two \cite{Du2005,Zhang2007}; this is not, however, a true example of the controlled-transfer gate approach as it does not prepare states with the highest possible purity.

\section{Implementations}
\begin{figure*}
\includegraphics[scale=1]{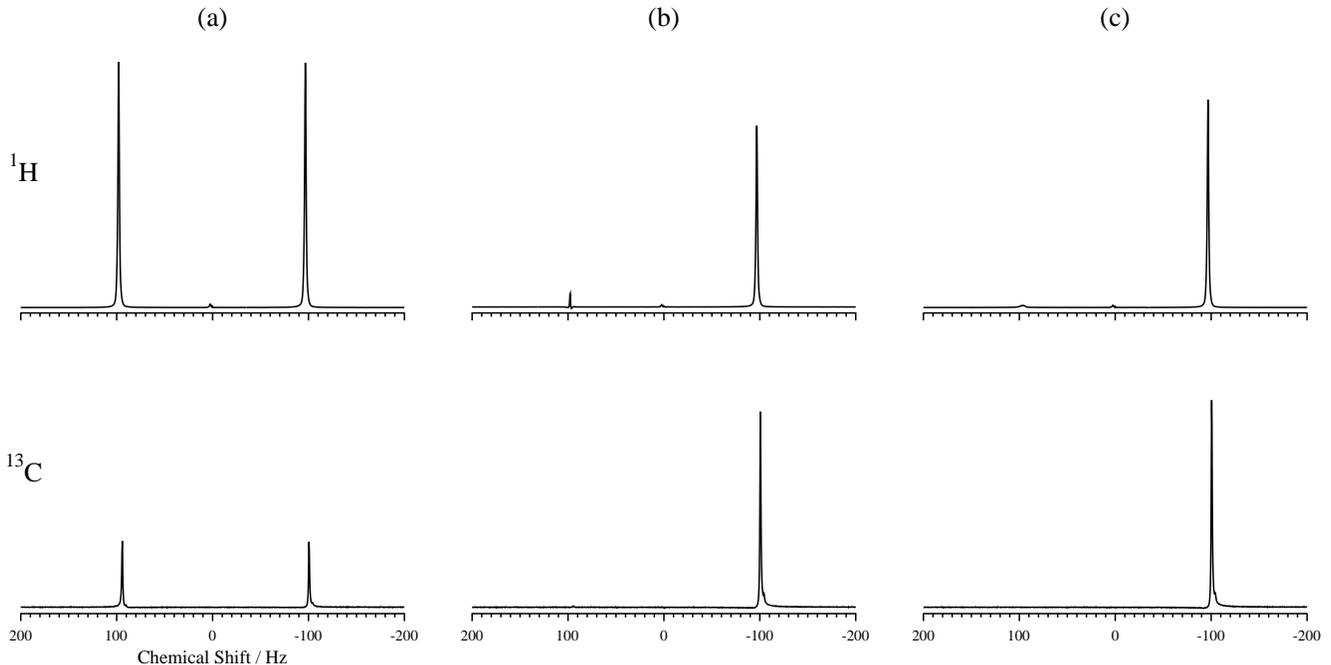}
\caption{Experimental demonstration of pseudo-pure states prepared using a conventional approach and our new approach.  The left hand spectra (a) show thermal states, while the central spectra (b) show pseudo-pure states prepared using a conventional approach \cite{Pravia1999}, and the right hand spectra (c) show pseudo-pure states prepared using our new approach.  The small signal visible at the centre of the \nuc{1}{H} spectrum arises from an impurity (unlabelled formate).  As absolute intensities are not meaningful in NMR spectra, the \nuc{1}{H} and \nuc{13}{C} spectra in (c) were normalised to have the same height \cite{Pravia1999}, and all other spectra plotted on the same vertical scale.}\label{Fig:expts}
\end{figure*}
Controlled-rotation gates can in principle be implemented as a pair of Hadamard gates applied to the target spin separated by a controlled-phase-shift gate.  Hadamard gates can be replaced by pseudo-Hadamard gates as usual \cite{Jones2001a}, while controlled-phase-shift gates correspond to a combination of evolution under the spin--spin coupling for some appropriate time, with undesired interactions being suppressed by spin echoes \cite{Jones1999}, and local $z$ rotations.  If desired, extraneous $z$ rotations and couplings can be absorbed into the spin reference frames \cite{Knill2000, Bowdrey2005}, and any such rotations remaining when the gradient pulse is applied can simply be dropped, potentially significantly simplifying the implementation.  Alternatively controlled-rotation gates can be implemented directly using methods such as \textsc{grape} sequences \cite{Schulte-Herbruggen2005,Khaneja2005}.

A controlled-transfer gate comprises a controlled-rotation gate followed by a crush gradient pulse to remove all off-diagonal terms from the density matrix.  Implementation of a single crush gradient is straightforward, but with two or more controlled-transfer gates it is necessary to consider the possibility of echoes from the interaction between two successive gradients separated by pulses \cite{EBWbook}.  These can be minimised by using gradients with different integrated strengths or along different axes, and will also be suppressed by decoherence and diffusion \cite{Cory1998}.

Alternatively the gradients can be replaced by temporal averaging, as applying a crush gradient is equivalent to averaging spectra acquired with and without the application of Z gates ($180^\circ_z$ rotations); when used with controlled-transfer gates it is only necessary to apply Z gates to the target qubit.  As before the Z gates can be implemented as frame rotations, and so this approach can be implemented by phase cycling \cite{EBWbook}, without changing the underlying pulse sequence.  This approach allows controlled-transfer gates to be used in experimental systems where spatial averaging is not available, but as the number of experiments required doubles with every gradient pulse replaced by Z gates the method is only practical for networks with small numbers of controlled-transfer gates.

Experiments were implemented using the two-qubit heteronuclear spin system (\nuc{1}{H} and \nuc{13}{C}) provided by dissolving \nuc{13}{C} labelled sodium formate in $\textrm{D}_2\textrm{O}$ \cite{Xiao2005}.  All experiments were performed at $20^\circ$C, on a Varian INOVA spectrometer with a nominal \nuc{1}{H} frequency of 600\,MHz.  Both nuclear spins were placed on resonance in their respective rotating frames, so the spin Hamiltonian contains only a scalar coupling term, with $\textrm{J}=194.7$\,Hz.  We chose to implement our controlled gates using sequences of pulses and delays, but all $z$-rotations (both desired rotations and extraneous rotations arising from chemical shift evolution) were absorbed into the reference frame.  The observed relaxation times were $\textrm{T}_1=5.5\,\textrm{s}$ and $\textrm{T}_2=0.61\,\textrm{s}$ for the \nuc{1}{H} qubit and $\textrm{T}_1=14.7\,\textrm{s}$ and $\textrm{T}_2=0.35\,\textrm{s}$ for the \nuc{13}{C} qubit.  All pulses used were naive rotations rather than composite pulses such as BB1 \cite{Cummins2003}.

The results, depicted in Fig.~\ref{Fig:expts}, show that this method can indeed be used to prepare pseudo-pure states.  Both \nuc{1}{H} and \nuc{13}{C} spectra are shown after applying a  $90^\circ$ excitation pulse to the spin subsequently observed; no signal is seen if no excitation pulse is applied or the excitation pulse is applied to the other spin (data not shown).  For a perfect pseudo-pure state each spectrum should show a single absorptive line on the right hand component of the doublet, with no signal in the other component, and the height of this line indicates the efficiency with which the pseudo-pure state has been prepared.  Any off-diagonal elements in the density matrix would be visible in spectra acquired without excitation pulses, or with these pulses applied to the wrong spin.  The absence of signal in these spectra (data not shown) indicates that the final crush gradient has been effective in removing off-diagonal terms.  Any deviations in the diagonal terms from the desired pattern for a pseudo-pure state, that is any imbalance in the three population terms supposedly equalised by the controlled-transfer gates, is visible as non-zero intensity on the left hand components of the doublets.  Pleasingly the error terms are visibly smaller with our new method, and the expected increase in signal size over the conventionally prepared state \cite{Pravia1999} is indeed seen.

\section{Asymptotic approaches}
Although the network described above, Eq.~\ref{eq:network2}, is relatively simple, it will in general require controlled-transfer gates with two different angles, and the initial state must be accurately known to determine these angles.  It is useful to consider what can be achieved with simpler networks such that $\theta_1=\theta_2=\theta$.  We will initially confine our discussion to homonuclear systems beginning at thermal equilibrium, and consider the simplest case, $\theta=\pi/2$.

With this restriction it is not possible to prepare a pseudo-pure state perfectly with a finite number of gates, and so it is necessary to have some criterion to measure the quality of approximate pseudo-pure states.  Conventional fidelity definitions are not appropriate, as these depend only on the population of the ground state \cite{Anwar2006} and are not affected by controlled-transfer gates.  Instead we use the square root of the $\chi^2$ difference between the diagonal density matrix $\rho$ and that of the desired pseudo-pure state $\rho_0$, divided by the corresponding value for the initial thermal state $\rho_i$
\begin{equation}
\epsilon(\rho)=\sqrt{\frac{\textrm{tr}[(\rho-\rho_0)^2]}{\textrm{tr}[(\rho_{i}-\rho_0)^2]}}=\frac{||\rho-\rho_0||_F}{||\rho_i-\rho_0||_F},
\end{equation}
where $||M||_F$ is the Frobenius norm of $M$, and seek the state with the lowest value of $\epsilon$.% (although in practice it may be easier to minimise $\epsilon^2$).

Choosing $\theta=\pi/2$ gives $\epsilon=1/4$, which is only a modest improvement on the initial value of $\epsilon=1$.  However applying the same network again drives the state closer to a pseudo-pure state, and applying the same network $r$ times
\begin{equation}
\mbox{
\Qcircuit @C=1em @R=.7em {
& \ctrl{1} & \sgate{\pi/2} & \qw \\
& \sgate{\pi/2}& \ctrl{-1} & \qw
\gategroup{1}{1}{2}{1}{.7em}{(}
\gategroup{1}{4}{2}{4}{.7em}{)}
}}\;{r}\label{eq:networka}
\end{equation}
gives $\epsilon=1/4^r$, so that repeated applications of the network drive the state asymptotically towards a pseudo-pure state.

For any given initial state it is possible to choose a value of $\theta$ which gives more rapid convergence to the pseudo-pure state; in general such values must be sought by numerical methods.  Some results for a homonuclear two spin system are given in table~\ref{tab:theta2}, showing that more rapid convergence is possible, but these values depend both on the choice of initial state and the number of times the network is applied, removing the intrinsic robustness of the asymptotic approach.
\begin{table}[t]
\begin{tabular}{|l|l|l|l|}
\hline
$r$&$\epsilon_{90}$&$\epsilon_\textit{opt}$&$\theta_\textit{opt}$\\\hline
1&0.2500&0.1437&$77.8^\circ$\\
2&0.0625&0.0294&$99.9^\circ$\\
3&0.0156&0.0008&$96.0^\circ$\\
4&0.0039&$4\times10^{-5}$&$95.0^\circ$\\
5&0.0010&$3\times10^{-6}$&$94.7^\circ$\\
\hline
\end{tabular}
\caption{The effectiveness of asymptotic preparation of a pseudo-pure state in a homonuclear two spin system, comparing optimal choices of the transfer angle $\theta$ with the naive choice of $90^\circ$.}
\label{tab:theta2}
\end{table}

A further advantage of this approach is that it works for \textit{any} initial state; although the exact results will depend on the choice of initial state the network will drive any state towards the desired pseudo-pure state.  Thus the same network can be used for homonuclear and heteronuclear spin systems, and for spin systems starting in non-thermal states.  In heteronuclear systems the state will converge more rapidly towards a pseudo-pure state if the first gate is applied with the more highly polarised spin as the target.  For any two spin heteronuclear system the optimal angle for the first gate lies within $20^\circ$ of the naive choice of $90^\circ$, and so convergence is quite rapid.

\begin{figure*}
\includegraphics[scale=1]{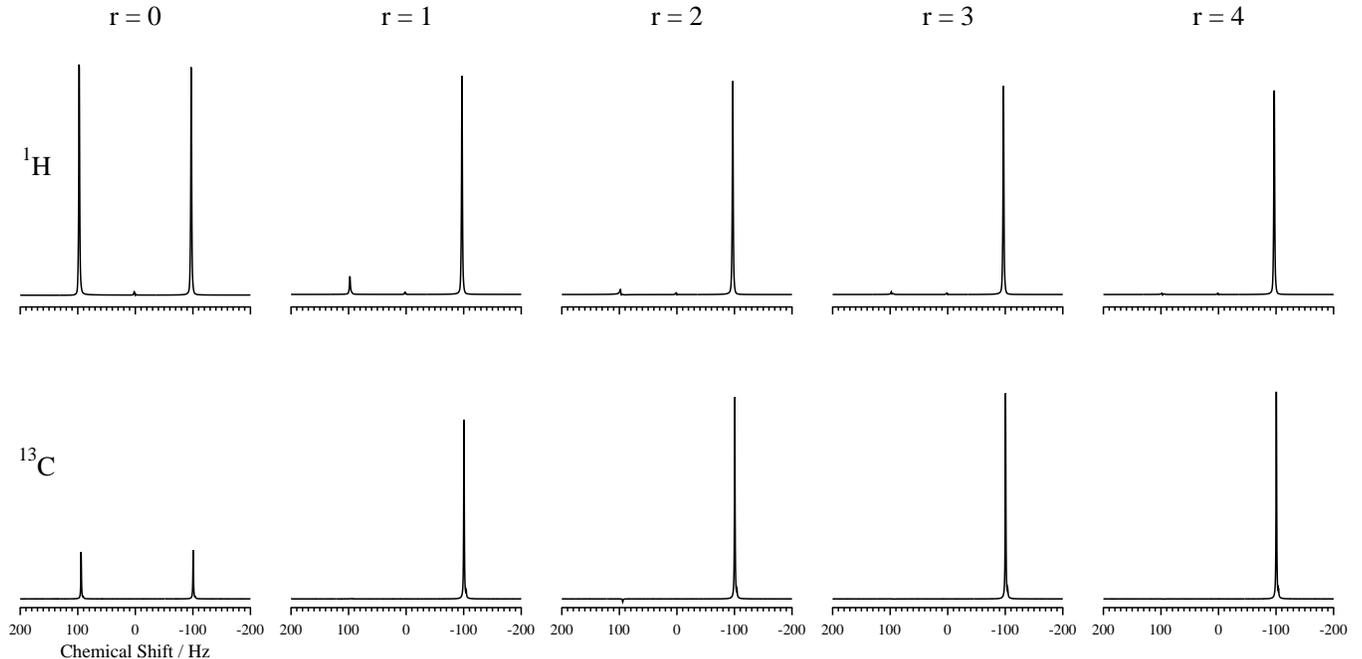}
\caption{Experimental demonstration of asymptotic preparation of pseudo-pure states.  Spectra are shown for between $r=0$ and $r=4$ rounds of the preparation sequence Eq.~\ref{eq:networka}.  In this case the spectra were acquired with an delay between experiments too short to allow complete relaxation, so the case $r=0$ is not simply the thermal spectrum, though it is broadly similar in form.  Pseudo-pure states prepared with this approach are essentially perfect for $r\ge3$.
%For more details see Fig.~\ref{Fig:expts}.
}\label{Fig:expts2}
\end{figure*}

The asymptotic network was implemented using the same heteronuclear spin system as before.  Fig.~\ref{Fig:expts2} shows spectra with between $r=0$ and $r=4$ rounds of the preparation sequence Eq.~\ref{eq:networka}, showing that the state converges to a pseudo-pure state around $r=3$.  These spectra were acquired with a total delay of 18\,s between individual scans, only slightly longer than the $\textrm{T}_1$ time of the slowly relaxing \nuc{13}{C} spin, so the initial state for these experiments was not the thermal equilibrium state.  As expected the asymptotic network is robust against changes in the initial state.

\section{Effects of decoherence and relaxation}
So far we have assumed that the controlled-transfer gates will be implemented perfectly, which is unlikely to be the case in a real system.  Systematic errors can be largely suppressed using composite pulses \cite{Cummins2003} or optimal control techniques such as \textsc{grape} \cite{Schulte-Herbruggen2005,Khaneja2005}, but decoherence and other forms of relaxation will provide a fundamental limit.  In the context of NMR it is useful to distinguish between $\textrm{T}_2$ processes, which cause coherent superpositions to decohere, and $\textrm{T}_1$ relaxation which acts to restore spin state populations to their equilibrium values.  Typically $\textrm{T}_2$ is significantly shorter than $\textrm{T}_1$, and so might be expected to dominate, but $\textrm{T}_2$ processes during a controlled-transfer gate will only affect the target spin, while $\textrm{T}_1$ processes will affect all the spins in the system.

The effect of $\textrm{T}_1$ processes in returning the spin state to the equilibrium populations can be modelled using generalized amplitude damping \cite{Vandersypen2001}, and it would in principle be possible to largely correct for these effects by subtly changing the angles in the controlled-transfer gates to compensate.  A simpler approach, however, is to use asymptotic sequences: as these drive the spin system towards the desired state, whatever the initial state, they will automatically compensate for small amounts of relaxation.  This analysis ignores the contribution of relaxation to the decoherence of off-diagonal elements during the implementation of a controlled-transfer gate, but these effects are indistinguishable from the pure decoherence effects considered below.

$\textrm{T}_2$ processes can be modelled in a similar way, and their effect is simply to drive the spin system towards the maximally mixed state.  As such they are mostly visible as a reduction in signal size, rather than errors in the state prepared.  This may be a problem, particularly in longer sequences such as those used with asymptotic approaches.  In our experiments, however, the effects of both $\textrm{T}_1$ and $\textrm{T}_2$ processes are small, as the relatively large J-coupling leads to comparatively short gate times.

\section{Larger spin systems}
Equivalent networks can be found in larger systems; in this case we will only consider homonuclear spin systems.  With three qubits there are six different controlled-transfer gates, leading to a much larger number of possible networks.  For a network of $p$ gates there are $6^p$ possible networks, each member of which has $p$ different transfer angles, $\theta_j$.  This large group can be immediately cut down in three ways.  Firstly the networks can be divided up into groups of six, with the members of each group differing only in the labelling of the qubits.  This is easily handled in a three qubit system by requiring the first gate to be controlled by qubit one and to target qubit two, implicitly labelling all three qubits.  Secondly, there is no point in applying the same gate twice in succession, as such gate pairs can always be replaced by a single gate with a different angle, and so there are only five sensible choices for each gate after the first one. Finally, to end in a pseudo-pure state, with all populations other than the ground state equal, the last gate must normally have a transfer angle of $\theta_p=\pi/2$.  This requirement can only be avoided if the network produces a pseudo-pure state after the first $p-1$ gates; in this case, since pseudo-pure states are invariant under all controlled-transfer gates, the final gate can have any transfer angle.  Even in this case, however, the network will work with $\theta_p=\pi/2$, so it is reasonable to impose this as a requirement.

A numerical search can then be used to minimise the error term $\epsilon$ for each of the $5^{p-1}$ networks (five choices for each gate except the first) over the $p-1$ variable transfer angles (the angles of all gates except the last).  A brute force search is perfectly practical for small values of $p$.  As before we assume that the spin system is homonuclear.  The first perfect solutions ($\epsilon=0$ within rounding errors) are found for networks of five gates, where 104 of the 625 distinct networks permit essentially perfect pseudo-pure states to be prepared.  With networks of only four gates, the best that can be achieved is $\epsilon\approx0.0391$, while with networks of six gates 1268 of the 3125 distinct networks, including all 120 networks containing one copy of each of the six gates, permit perfect pseudo-pure states to be prepared.

With systems of three or more qubits it is likely that it will not be equally easy to implement controlled-transfer gates between all pairs of qubits.  An important limiting case is when the qubits form a linear chain, with each qubit only having couplings to its immediate neighbours, so that nearest-neighbour gates are simpler to implement than long range gates.  This is not a fundamental problem, as indirectly coupled gates \cite{Collins2000} can be used to implement long range gates with only moderate overhead, but it is useful to investigate whether pseudo-pure states can be prepared using networks containing only nearest-neighbour controlled-transfer gates.

For a three qubit system, with four possible nearest-neighbour gates, it is once again relatively simple to investigate this numerically.  The imposition of a chain structure means that it is no longer possible to treat the three qubits as identical, but a saving of a factor of two can be made by noting the symmetry between the two possible orientations of the chain (qubit 1 first, and qubit 3 first), so in this case there are $2\times3^{p-1}$ basic networks.  Once again the first solutions are found for the case $p=5$, where two of the 162 distinct networks permit perfect pseudo-pure states to be prepared.  The explicit networks are
\begin{equation}
\mbox{
\Qcircuit @C=1em @R=.7em {
&\sgate{\theta_1}&\qw        &\ctrl{1}     &\qw             &\qw          &\qw\\
&\ctrl{-1}       &\ctrl{1}   &\sgate{\pi/2}&\ctrl{1}        &\sgate{\pi/2}&\qw\\
&\qw             &\sgate{\pi}&\qw          &\sgate{\theta_2}&\ctrl{-1}    &\qw
}}\label{eq:network3a}
\end{equation}
and
\begin{equation}
\mbox{
\Qcircuit @C=1em @R=.7em {
&\sgate{\pi}&\qw             &\qw          &\sgate{\theta_2}&\ctrl{1}     &\qw\\
&\ctrl{-1}  &\ctrl{1}        &\sgate{\pi/2}&\ctrl{-1}       &\sgate{\pi/2}&\qw\\
&\qw        &\sgate{\theta_1}&\ctrl{-1}    &\qw             &\qw          &\qw
}}\label{eq:network3b}
\end{equation}
with the angles $\theta_1=\arccos(-1/7)\approx98.2^\circ$ and $\theta_2=\arccos(-5/7)\approx135.6^\circ$. These solutions (together with the equivalent pair obtained by swapping qubits one and three) appear to be the simplest networks possible; in particular, as previously noted, it is not possible to prepare a pseudo-pure state of a three qubit system using only four controlled-transfer gates even if long-range gates are permitted.

\section{Asymptotic approaches}
\begin{figure}[t]
{\includegraphics[width=85mm]{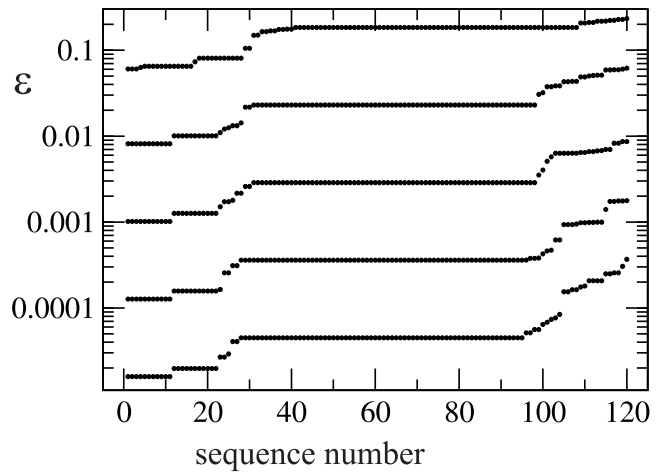}}
\caption{The performance of asymptotic networks in a three qubit system.  The error term $\epsilon$ is shown for each of the 120 distinct networks of six different gates applied to a homonuclear system.  The networks are plotted ordered from lowest to highest error, and the five ``lines'' correspond to applying the networks from one to five times.}
\label{fig:ass}
\end{figure}

The asymptotic approach can also be used in these larger spin systems.  We begin by considering the 120 distinct networks containing one copy of each of the six gates, where each gate is assumed to use the same transfer-angle, $\theta=\pi/2$.  The majority of these networks, 68, give the same error, $\epsilon\approx0.1840$, but 40 networks give lower errors than this, and three achieve the lowest value of $\epsilon\approx0.0605$, while 12 networks give higher errors, with the worst case error, $\epsilon\approx0.2332$, occurring for a single network.  As usual these results can be improved by applying the network repeatedly, and for the main group of 68 networks the error is given by
\begin{equation}
\epsilon(r)=\frac{\sqrt{13/6}}{8^r}.\label{eq:eps68}
\end{equation}
The behaviour of the other networks is more complicated, and is shown in outline in Fig.~\ref{fig:ass}.
As before, the behaviour of these networks can be improved by allowing $\theta$ to vary.  However initial studies suggest that, unlike the two-qubit case, only modest improvements are possible, and so we do not pursue this point further.

Next we turn to the use of asymptotic networks in linear chains where only nearest-neighbour gates are permitted, beginning with the twelve distinct networks of four gates.  None of these are effective at forming a pseudo-pure state, either for the case $\theta=\pi/2$ or when $\theta$ is optimised.  Applying the networks repeatedly helps a little, but is not as effective as was seen in the two-qubit case.  A likely explanation for this can be seen by examining the networks (\ref{eq:network3a}) and (\ref{eq:network3b}): both of these contain controlled-transfer gates with angles of $\pi$, which act to invert population differences within the system, rather than average them out, and this inversion cannot be well approximated by any set of averaging gates.

This can be addressed by relaxing the conditions slightly, permitting the use of two angles, namely $\pi/2$ and $\pi$, and searching over the 12 distinct patterns of four gates with 16 different sets of transfer angles, giving 192 possible networks.  The optimal sequence was found to be
\begin{equation}
\mbox{
\Qcircuit @C=1em @R=.7em {
&\sgate{\pi/2}&\qw          &\qw        &\ctrl{1}     &\qw\\
&\ctrl{-1}    &\sgate{\pi/2}&\ctrl{1}   &\sgate{\pi/2}&\qw\\
&\qw          &\ctrl{-1}    &\sgate{\pi}&\qw          &\qw
\gategroup{1}{1}{3}{1}{.7em}{(}
\gategroup{1}{6}{3}{6}{.7em}{)}
}}\;{r}\label{eq:network3ass}
\end{equation}
for which the error term is given by equation~\ref{eq:eps68}; thus this four gate sequence can do as well as the majority of six gate asymptotic networks using only angles of $\pi/2$, but it does not perform as well as the best ones.

Finally we consider the case of networks made up solely of nearest neighbour gates with transfer angles of $\pi/2$ and $\pi$, but relax the requirement for repeated application of a simple pattern of gates, seeking instead the optimal network with some total number of gates, $p$.  As before such networks occur in mirror symmetric pairs, should not contain repeated gates, and should end with a gate with a transfer angle of $\pi/2$, resulting in a final list of $4\times6^{p-2}\times3$ distinct useful networks, and we have explored these numerically for values of $p$ between four and eight.  As expected increasing $p$ allows better networks to be located, but the best networks for $p=4$ and $p=8$ are no better than the asymptotic networks found previously.

\section{Conclusions}
Controlled-transfer gates provide a simple and effective way of preparing pseudo-pure states in small spin systems, and experimental implementations confirm that as expected these networks give a larger signal intensity than conventional approaches.  While optimised networks provide the simplest approach when the initial state is well understood the less-efficient asymptotic approach is more robust.  In three-spin systems it is possible to find both optimised and reasonable asymptotic networks if all gates are allowed, but if the spin system is assumed to have a chain topology only the optimised approach leads to reasonable networks.  Initial studies (data not shown) suggest that similar results will be obtained in larger spin systems.

\begin{acknowledgments}
We thank the UK EPSRC and BBSRC for financial support.
\end{acknowledgments}

\bibliography{ctransfer}

\end{document}